\documentclass[fleqn,usenatbib]{mnras}
\usepackage[normalem]{ulem}
\usepackage{amssymb}
\usepackage{varwidth}
\usepackage{graphicx}                   
\usepackage{float}
\usepackage{amsmath}
\usepackage{amssymb}
\usepackage{multirow}
\usepackage{caption}
\usepackage{bm}
\usepackage{color}
\usepackage{ulem}
\usepackage[figuresright]{rotating}
\usepackage{appendix}
\usepackage{adjustbox}
\usepackage{txfonts}
\usepackage[justification=centering]{caption}
\usepackage[T1]{fontenc}

\DeclareRobustCommand{\VAN}[3]{#2}
\let\VANthebibliography\thebibliography
\def\thebibliography{\DeclareRobustCommand{\VAN}[3]{##3}\VANthebibliography}

\usepackage{graphicx}   
\usepackage{amsmath}    
\usepackage{amssymb}    

\title[]{Evidence of the gamma-ray counterpart from nova FM Cir with $Fermi$-LAT}

\author[H.H. Wang et al.]{
H.H. Wang$^{1}$\thanks{E-mail: wanghh33@mail.sysu.edu.cn},
H.D. Yan$^{2}$\thanks{E-mail: yanhd125@alumni.hust.edu.cn},
L.C.-C. Lin$^{3}$\thanks{E-mail: lupin@phys.ncku.edu.tw},
J. Takata$^{2}$\thanks{E-mail: takata@hust.edu.cn}, P.-H. T. Tam$^{1}$\thanks{E-mail:
tanbxuan@mail.sysu.edu.cn}
\\
$^{1}$School of Physics and Astronomy, Sun Yat-sen University, Zhuhai 519000, China\\
$^{2}$Department of Astronomy, School of Physics, Huazhong University of Science and Technology, Wuhan 430074, China\\
$^{3}$Department of Physics, National Cheng Kung University, Tainan 701401, Taiwan}

\begin{document}
\maketitle

\begin{abstract}
  We report the analysis results of X-ray and $\gamma$-ray data  of  the nova FM Cir taken by
  Swift and $Fermi$-LAT. The $\gamma$-ray emission from FM Cir can be identified with a significance level of $\sim$ 3$\sigma$ within $\sim$ 40 days after the nova eruption (2018 January 19) while we bin the light curve per day. The significance can further exceed 4$\sigma$ confidence level if we accumulate longer time (i.e., 20 days) to bin the light curve.
  The $\gamma$-ray counterpart could  be identified with a Test Statistic (TS) above 4 until 
  $\sim 180$~days after the eruption. The duration of the gamma-ray detection was longer than those reported in the previous studies of the other novae detected in the GeV range.
  The significant X-ray emission was observed after the gamma-ray flux level fell below
  the sensitivity of $Fermi$-LAT.  The hardness ratio of the X-ray emission decreased rapidly with time, and
  the spectra were dominated by blackbody radiation from the hot white dwarf. Except for the longer duration of the $\gamma$-ray emission, the multi-wavelength properties of FM Cir closely resemble those of other novae detected in the GeV range. 

\end{abstract}
\begin{keywords}
stars: individual: FM Cir-stars: novae, cataclysmic variables – stars: white dwarfs. 
\end{keywords}

\section{Introduction}
Classical novae are thermonuclear eruptions that occur in binary systems, where a white dwarf accretes matter from its binary companion. The energy release resulting from the thermonuclear eruption leads to a dramatic expansion and ejection
  of the accretion envelope. It has been observed that the ejected matter expands into the surrounding environment at a speed of hundreds to
  thousands of ${\rm km~s^{-1}}$  \citep{Gallagher1978}. Multi-wavelength observations in the past decade have revealed that novae are transient sources in the broadband energy bands from radio to gamma-rays~\citep{Chomiuk2021,Acciari2022NatAs,HESS2022}. It is argued that the shock is formed due to the collisions of the multiple ejecta (internal shock) or the interaction between the ejecta and preexisting medium surrounding the binary \citep{Della2020A&AR,Aydi2020ApJ,Chomiuk2021}.

$Fermi$ Large Area Telescope ($Fermi$-LAT) had discovered GeV emissions from  19~novae  and potential emissions from 6 sources\footnote{\url{https://asd.gsfc.nasa.gov/Koji.Mukai/novae/latnovae.html}}. \cite{Abdo2010} reported the first nova detected in the GeV bands, V407 Cyg, which is the binary system composed of a white dwarf and a red giant companion. It is interpreted that the shock was formed as a result of the collision between the nova ejecta and the dense wind from the red giant companion. Another nova, RS~Oph, with a red giant companion, was detected in TeV energy bands, and its TeV light curve shows a temporal evolution similar to the GeV emission,  indicating  GeV and TeV emissions from RS~Oph  have 
a common origin each other~\citep{HESS2022}.  Most of the {novae detected in the GeV range} are classified as classical nova and have a main-sequence star as the companion. For such a system, it has been argued that the $\gamma$-ray  emission originates from the internal shock, because the stellar wind from the main-sequence star is too weak to create a strong shock~\citep{Chomiuk2021}. {It has been  suggested that a shock of the ejected matter by nova eruption
  accelerates particles to relativistic energy, leading to the production of gamma-rays through leptonic and/or hadronic processes~\citep[e.g.][]{vurm2018, Chomiuk2021}.
  In the leptonic scenario, for example, the accelerated electrons produce the  $\gamma$-rays via bremsstrahlung and/or inverse Compton scattering processes.
  In the hadronic scenario, the pion decay produces the majority of the gamma-rays, and the secondary electron/positron pairs also contribute to the emission
   through the bremsstrahlung or inverse-Compton scattering processes.}

Observed X-ray emission of novae is usually characterized by thermal emission of the hot white dwarf and/or the shocked matter~\citep{Orio2001A&A,Mukai2008,Chomiuk2014Natur}. The soft X-ray emission with an effective temperature of $<0.1$~keV can reach a luminosity of $L_X>10^{36}~{\rm erg~s^{-1}}$, and it is considered that the so-called super-soft emission originates from a hot white dwarf sustained by residual nuclear burning. The X-ray emission from the majority of the {novae detected in the GeV range} appears after the $\gamma$-ray flux falls below the detector sensitivity of the $Femi$-LAT. This is interpreted that it in the earlier epoch is absorbed by the ejecta and is reprocessed into the emission in lower energy bands~\citep{Metzger2014,Li2017,Aydi2020}.
As the ejected material spreads out, the environment becomes optically thin and makes  the soft X-ray emission from the hot white dwarf visible~\citep{Page2020AdSpR}.

A naked-eye nova, FM Cir (also known as Nova Cir 2018 or PNV~J13532700-6725110), was discovered by John Seach on 2018 January 19~\footnote{\url{https://www.aavso.org/vsx/index.php?view=detail.top\&oid=555615}}, in the constellation of Circinus. FM Cir has been classified as a classical Fe II nova through optical spectroscopy \citep{Strader2018}, and  the binary system has a 3.4898-day of orbital period \citep{Schaefer2021}. FM Cir was extensively monitored optically by the AAVSO (American Association of Variable Star Observers) group, and its light curve exhibited multiple peaks and an absorption of the dust~\citep{Molaro2020MNRAS}.

In this paper, we report analysis results of the   GeV and X-ray observations for  FM Cir, and we find evidene that  
the duration of the $\gamma$-ray detection is longer to those in other {novae detected in the GeV range}. In Section 2, we describe the data analysis. The results from the $\gamma$-ray and X-ray data analysis are presented in Section 3, along with a comparison of nova FM Cir with other classical novae. We make a conclusion in Section 4.

\section{data reduction}

\subsection{LAT data analysis}
We performed a binned analysis using the standard $Fermi$-LAT ScienceTool package, which is available from the $Fermi$ Science Support Center\footnote{\url{https://fermi.gsfc.nasa.gov/ssc/data/access/lat/}}.
The data for the fourth Fermi-LAT catalog (4FGL DR4, \verb|gll_psc_v32.fit|) were taken during the period August 2008 to August 2022 covering 14 years~\citep{4fgl-dr4,4fgl-dr3}, which contains $\sim$ 7200 gamma-ray sources. To avoid Earth's limb contamination, we only included the events with zenith angles below 90 degrees. We limited our analysis to the events from the point source or Galactic diffuse class (\verb|event class = 128|) and used data from both the front and back sections of the tracker (\verb|evttype = 3|).
We constructed a background emission model that incorporates both the Galactic diffuse emission (\verb|gll_iem_v07|) and the isotropic diffuse emission (\verb|iso_P8R3_SOURCE_V3_v1|). A $\gamma$-ray emission model for the whole ROI was built using all sources in the fourth $Fermi$-LAT catalog~\citep{FERMI2020} located within $20^{\circ}$ of the nova, and the nova FM Cir is included in the model at the nova position of (R.A., decl.)=($13^{o}53^{'}27.61^{''},-67^{o}25^{'}00.9^{''}$).

\begin{figure}
    \centering
    \includegraphics[scale=0.54]{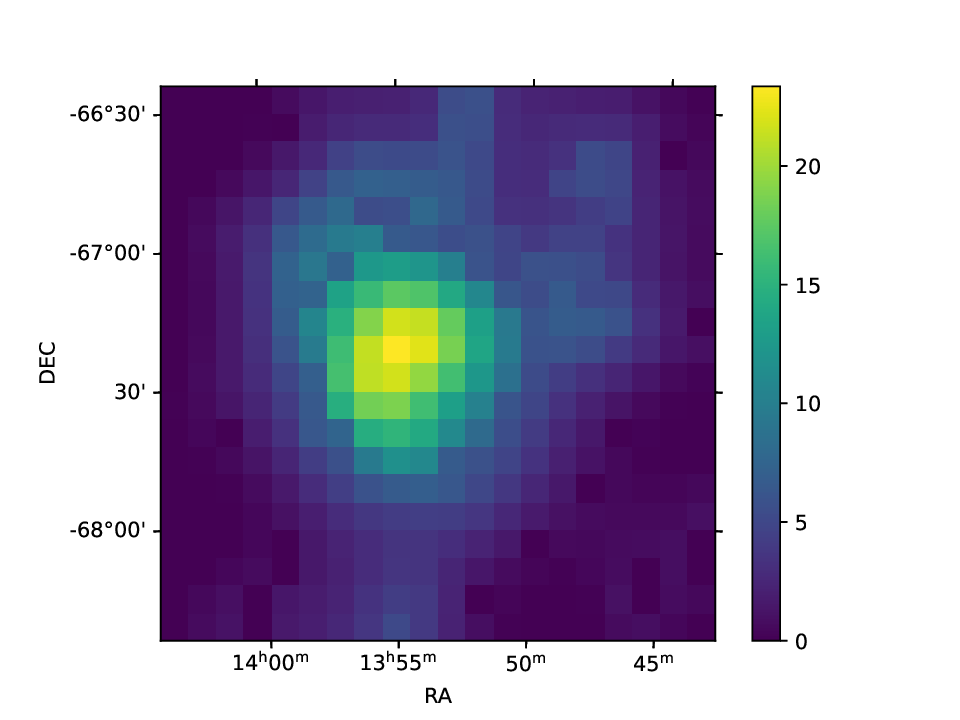}
    \caption{The TS map of nova FM Cir in 0.1-300~GeV from MJD~58137 to MJD~58178.}
    \label{map}
\end{figure}

\begin{figure}
\centering
\includegraphics[scale=0.30]{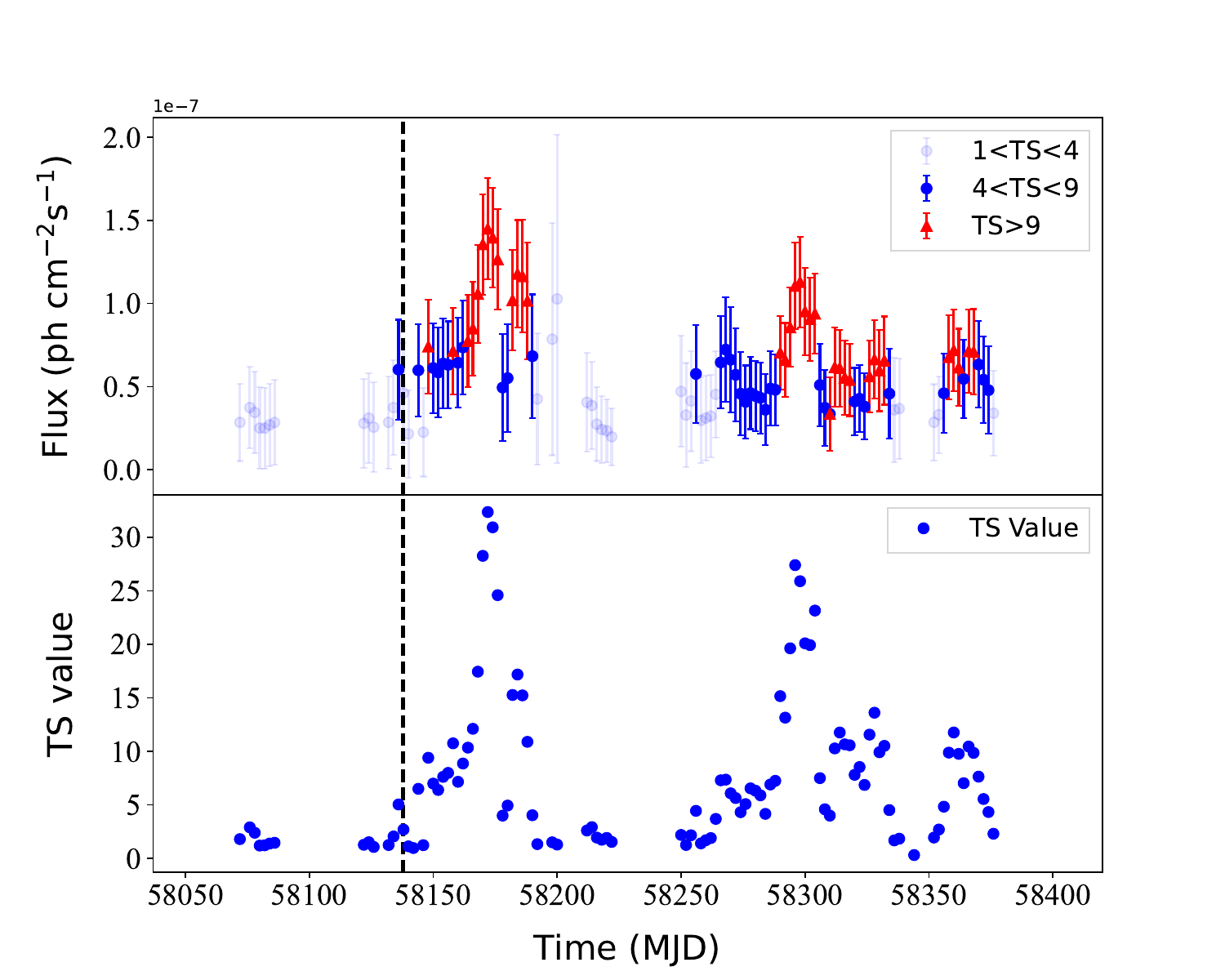}
\caption{\label{lcnova} The light curve of the Nova FM Cir in a sliding time window of 20 day length, the time window is shifted in 2 day steps. Upper panel: the evolution of flux above $3\sigma$~(red triangles) and the flux with TS value~(Test Statistic) in the range of 4 to 9~(deep blue dots), TS value small than 4~(light blue dots).
Bottom panel: the evolution of TS value. The vertical dash line shows the epoch of nova eruption in optical. }
\end{figure}

\begin{figure*}
    \centering
    \includegraphics[scale=0.45]{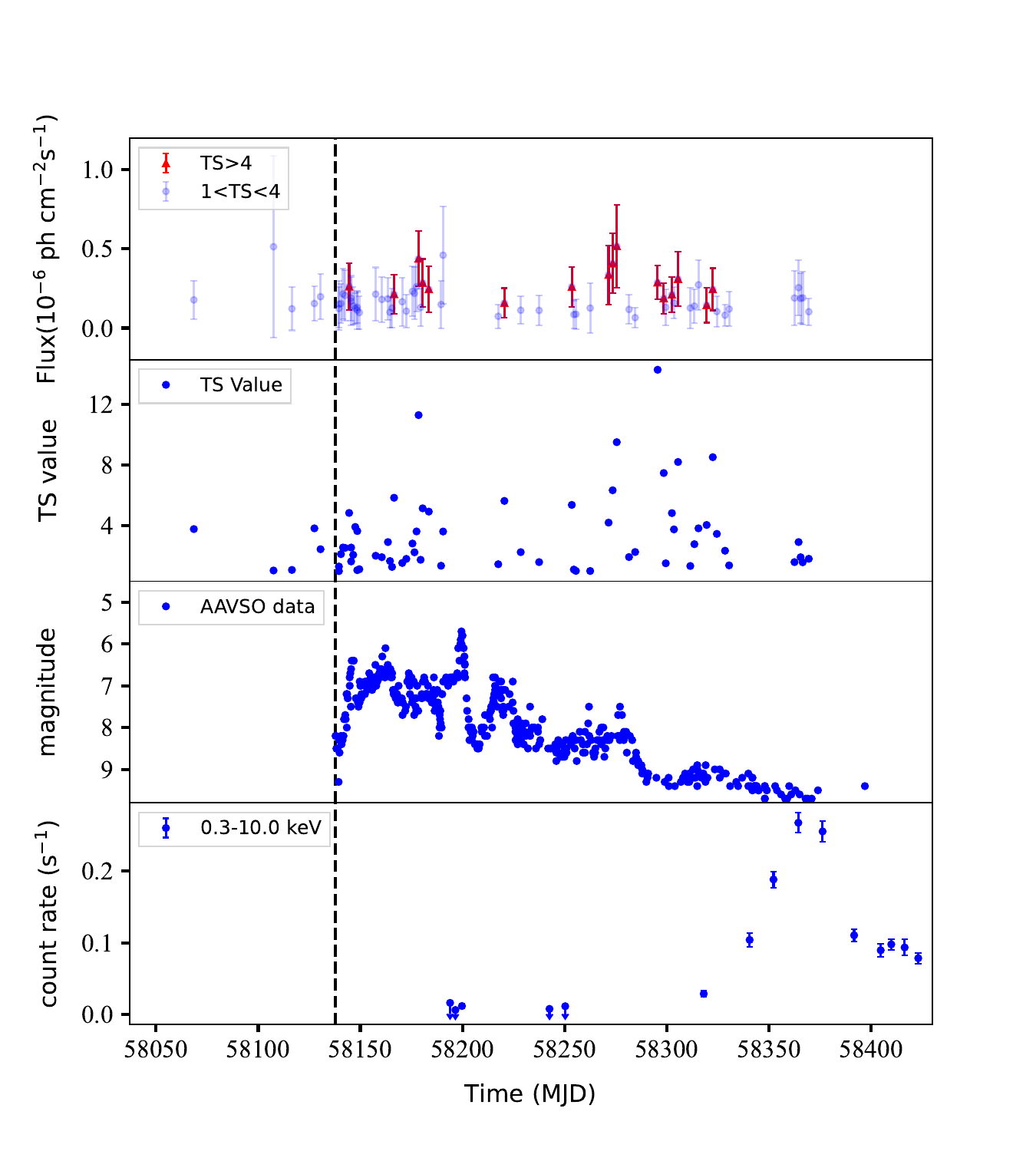}
    \caption{The photon flux with 1-day bin using $Fermi$-LAT data. The red triangles and the light blue dots in upper panel are the flux with the TS value large than 4~(corresponding to a detection significance of $\sim$ 2~$\sigma$) and 1~($\sim$1~$\sigma$), respectively. Second panel: the evolution of TS value. Third panel: the light curve from optical data in site of \url{http://aavso.org/lcg}. Fourth panel: the count rate of X-ray data which come from the $Swift$ telescope in 0.3-10.0~keV. The vertical dash line shows the epoch of nova eruption in optical.}
    \label{lighcurve-1d}
\end{figure*}

We utilized the Test Statistic (TS) map to scrutinize the morphology of the emission originating from the FM Cir source region (Figure~\ref{map}).
To examine the temporal evolution of the gamma-ray emission, we selected the $Fermi$-LAT data ($>100~$MeV) taken from MJD 58037, which is $\sim$100 days before the eruption (at 2018 January 19 = MJD~58137),
to MJD~58430, after which no GeV detection is confirmed.  We used \verb|gtlike| to search for the emission with a 20-day time window
in a two-day time step and plot the data points with the TS value above 1 (see Figure~\ref{lcnova}). We also created the daily light curve, which allows a more precise measurement for the epoch of the gamma-ray emission as shown in Figure~\ref{lighcurve-1d}. To generate the spectrum, we performed the likelihood analysis using the data obtained from MJD~58137 to MJD~58178, around the epoch of the optical peak and of the gamma-ray detection with a TS value of 9 (i.e., $>3~\sigma$).

\subsection{Swift data analysis}
\begin{figure}
    \includegraphics[scale=0.28,angle=270]{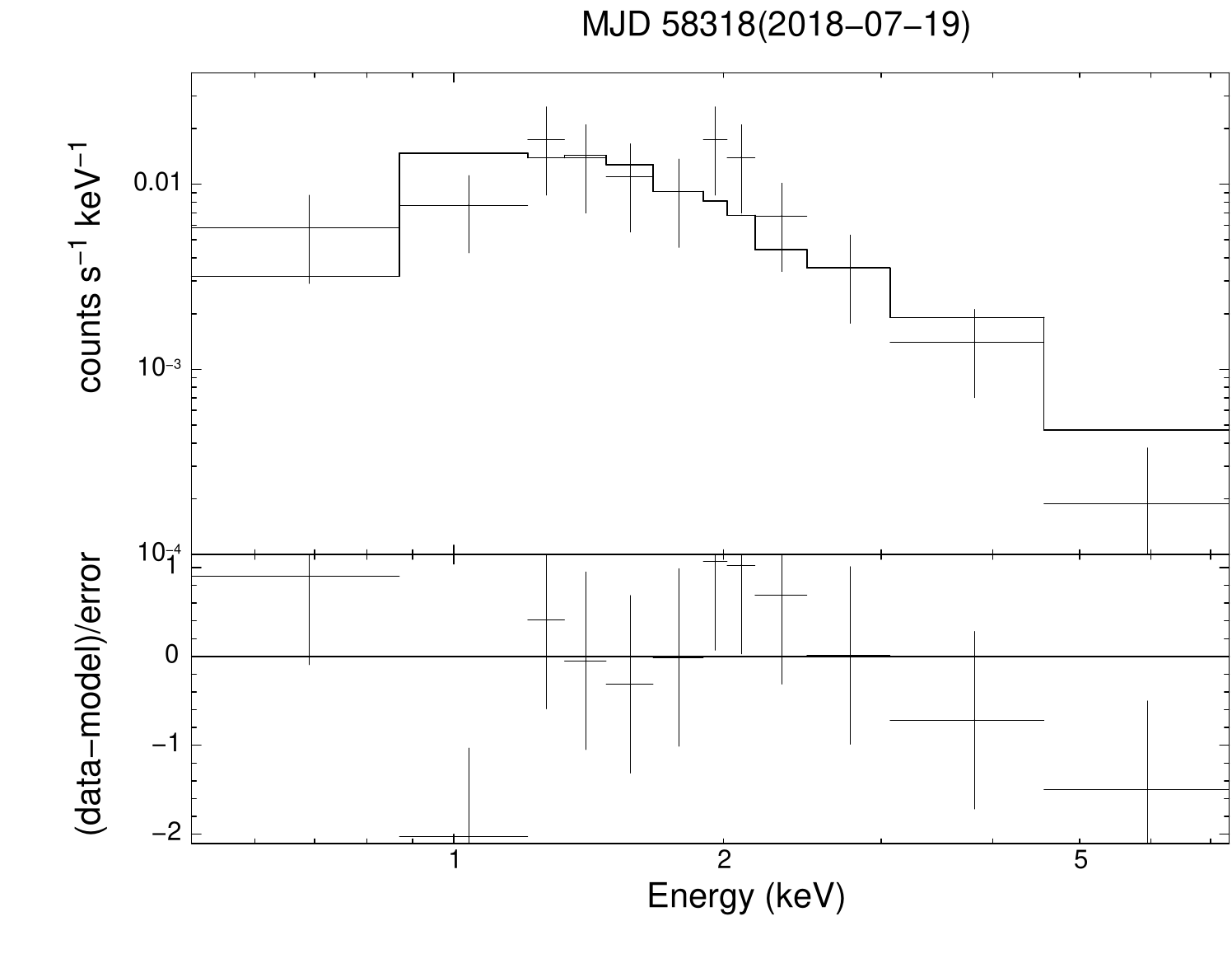}
    \includegraphics[scale=0.28,angle=270]{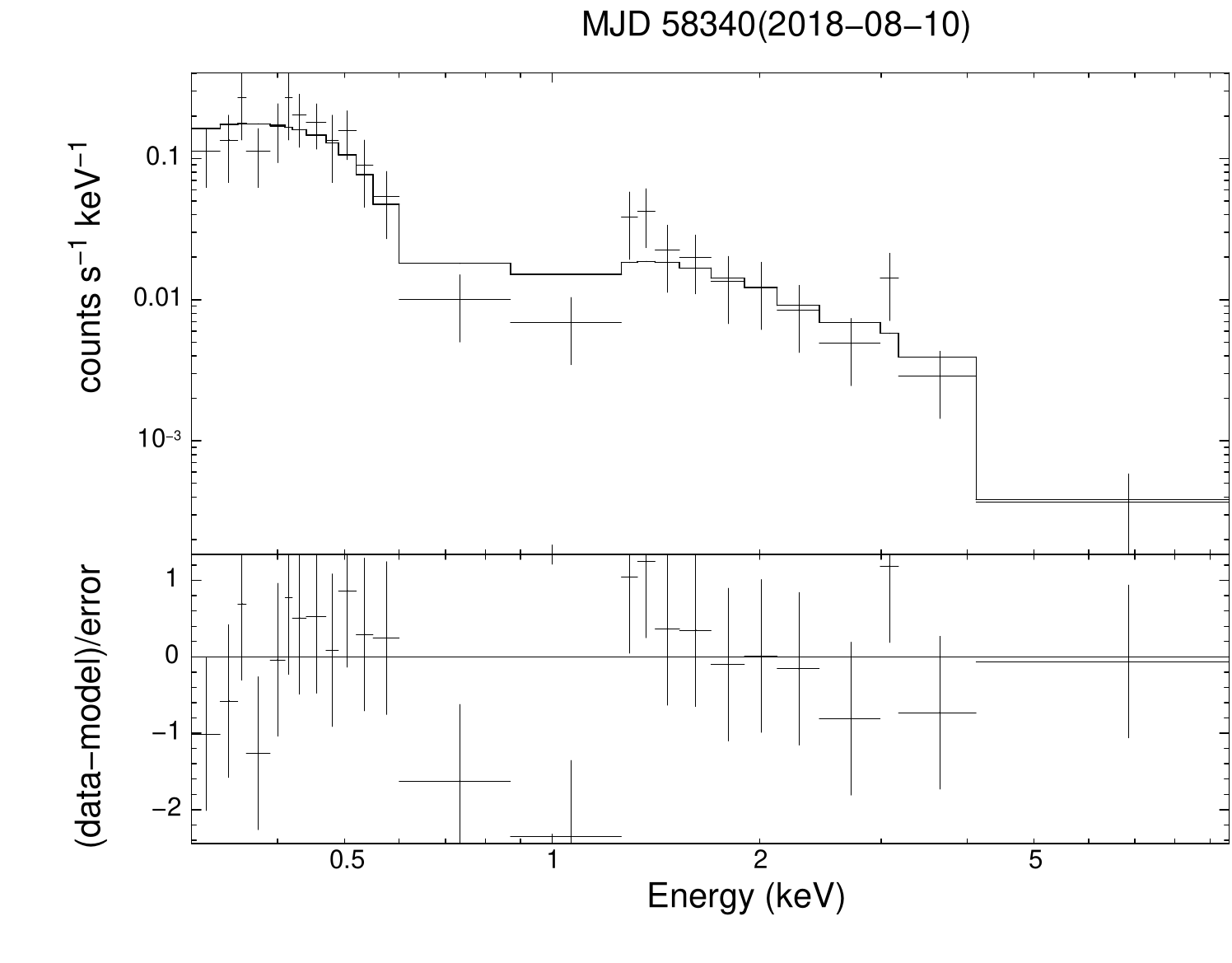}
    \caption{{X-ray  spectrum of nova FM Cir taken at  MJD 58318 (the first significant detection by the  Swift observation) and MJD 58340. All spectra taken at different epoch are presented in Figure~\ref{spectrum-X-ray-all}}.}
    \label{spectrum-X-ray}
\end{figure}

Swift had continuously monitored FM Cir from the discovery of the nova eruption until to $\sim$MJD~58425. We created the light curve and hardness ratio of the X-ray emission using the XRT web tool\footnote{\url{https://www.swift.ac.uk/user\_objects/}} \citep{Evans2007A&A,Evans2009MNRAS}. 
To investigate the spectral properties, we downloaded the archival data from HEASARC Browse\footnote{\url{https://heasarc.gsfc.nasa.gov/cgi-bin/W3Browse/w3browse.pl}} and performed 
the analysis with the HEASOFT version 6.31.1 and the  SWIFTDAS package with the updated calibration files. The clean event lists were obtained using the task  \verb|xrtpipeline| of the HEASOFT and extract the spectrum using \verb|Xselect|. We grouped the source spectra to ensure at least 1 count per spectral bin and fit the spectra using \verb|Xspec|. We employed the blackbody radiation and/or optically thin plasma emission to  fit the observed spectra.
{Figures~\ref{spectrum-X-ray} and~\ref{spectrum-X-ray-all} present the X-ray spectra taken at the different epoch. }
\begin{table*}
\centering
\caption{{The fitting parameters of X-ray spectra
    using the model of blackbody radiation and/or  optically thin thermal plasma emission (mekal).}}
\begin{tabular}{|c|c|c|c|c|c|c|c|}
    \hline
    \hline
ObsID&date(MJD) 
&N$_H$($10^{21}~cm^{-2}$) 
&$k_{B}T$$_{BB}$~(eV)
&F$^a_{BB}$
&$k_{B}T$$_{mekal}$~(keV) 
&F$^a_{mekal}$
&C-Stat/dof  \\
         \hline
00010625009&58318
&4.30$\pm$1.56&$-$&$-$&2.94$\pm$1.39&1.75$\pm$0.41&44/41
\\
00010625010&58340
&3.96$\pm$1.27&32.13$\pm$6.87&55.59
$\pm$11.03&4.68$\pm$2.66&0.29$\pm$0.07&69/73
\\
00010625011&58352 
&2.94 &28.76$\pm$1.41&72.44
$\pm$7.87&2.31$\pm$0.58&0.19
$\pm$0.041&71/87
\\
00010625012&58364
&2.94&27.33$\pm$0.89&141.25
$\pm$12.43&1.62$\pm$0.28& 0.12$\pm$0.031&88/75
\\
00010625013&58376
&2.94&29.90$\pm$1.05&104.71
$\pm$9.21&0.64$\pm$0.061&0.16$\pm$0.04&104/77
 \\
00010625014&58391
&2.94&26.68$\pm$1.49&64.56
$\pm$8.33&0.53$\pm$0.13&0.083$\pm$0.0324&40/51
\\
00010625015&58404
&2.94&38.87$\pm$3.91&15.13
$\pm$2.55&0.69$\pm$0.18&0.067$\pm$0.037&36/43
\\
00010625016&58409
&2.94&27.10$\pm$1.81&51.28$\pm$6.62&
0.76$\pm$0.26&0.074$\pm$0.028&47/49
\\
00010625017&58416
&2.94&31.84$\pm$5.06&26.91
$\pm$6.02&0.17$\pm$0.026&0.41$\pm$0.32&26/34
\\
00010625018&58423
&2.94&31.30$\pm$3.64&21.87
$\pm$2.38&0.18$\pm$0.036&0.19$\pm$0.12&42/44
\\

  \hline
    \end{tabular}
    \label{table-model}
~~~~~~~~~~~~~~~~~~~~~~~~~~~~
~~~~~~~~~~~~$^a$The unabsorbed flux is measured in 0.3-10.0 keV and recorded in units of 10$^{-11}$ erg cm$^{-2}$ s$^{-1}$.
\end{table*}

 \section{Results and discussion}
 \subsection{ GeV emission properties}
 Figure~\ref{map} illustrates the TS map of the nova FM Cir region in 0.1-300~GeV energy bands with the data taken at  MJD~58137-58178. The maximum TS value is around 22, which corresponds to a detection significance of larger than $4\sigma$~(i.e., $\sqrt{{\rm TS}}$ is about detection significance in $\sigma$).
Figure~\ref{lcnova} shows the light curve created with 20-day time wind.  We can find in the figure  
that TS value increased after nova eruption, reaching TS$\sim$ 30 at MJD 58175 and MJD 58300. In 
the daily light curve of Figure~\ref{lighcurve-1d}, we can also confirm the emission with $TS>9$ at MJD~58175 and MJD~58300.  The delay of the $\gamma$-ray peak from the epoch of the nova eruption is  one of the common properties among the novae detected in GeV range, and it may be due to the timescale for particle acceleration in the ejecta or $\gamma$-ray absorption by a dense ejecta~\citep{Ackermann2014,Metzger2015}. After the first peak occurred at around  MJD~58175, both TS values and flux decreased with time. As shown in Figures~\ref{lcnova} and~\ref{lighcurve-1d}, however, there is evidence of the second peak at around MJD~58300, which is about 125 days after the first peak. In the light curve of daily bins, the emission with TS >4 ($\sigma>2$) were confined until $\sim 180$~days after the nova eruption. 

We fitted  the spectrum obtained in MJD~58137-58178 (around the first peak) using  a power-law function with   an exponential cut-off of  
\begin{equation}
\centering
\label{eq2}
    \frac{dN}{d{E}} \propto E^{-\gamma_{1}}{\rm exp}\left[-\left(\frac{E}{E_c}\right)^{\gamma_{2}}\right],
\end{equation}
where we fixed to $\gamma_2=2/3$. We obtained a power-law index 
of $\gamma_1=1.88(2)$ and a cut-off energy of $E_{c}=1.00(5)$~GeV, which are similar to those of the novae detected in GeV bands ~\citep{Franckowiak2018,Chomiuk2021}. We obtained an  averaged energy flux of  
$F_{\gamma}=7.1(3)\times 10^{-12}~{\rm erg~cm^{-2}s^{-1}}$ in 0.1-300 GeV bands.  For the second peak in MJD~58280-58312, we fitted the  spectrum using
a pure power law function and obtained  a photon index of $2.12(2)$.  The  time averaged energy flux is  $F_{\gamma}=2.52(1)\times 10^{-12}~{\rm erg~cm^{-2}s^{-1}}$ in 0.1-300 GeV bands.

\begin{figure}
     \centering
     \includegraphics[scale=0.34]{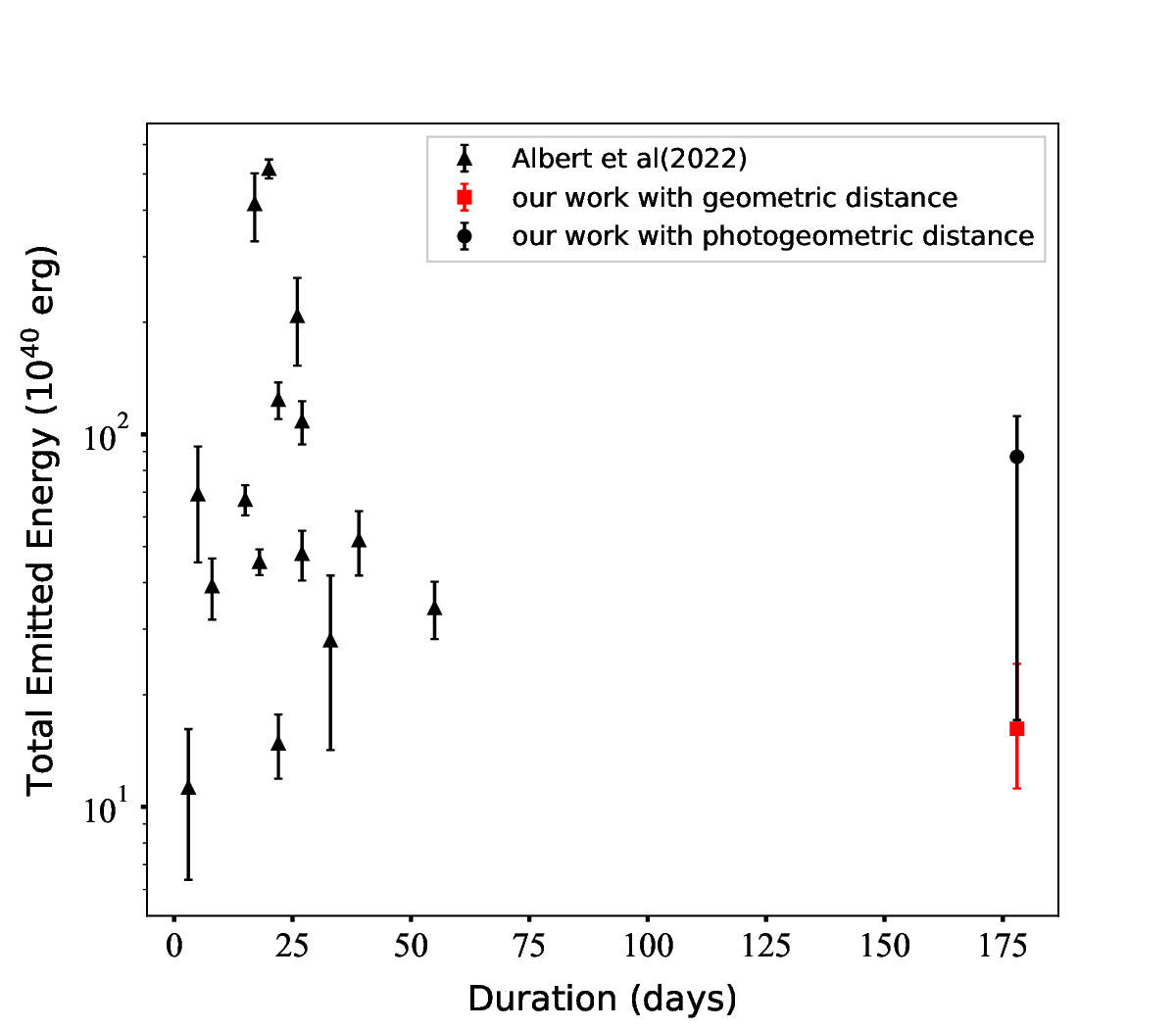}
     \caption{Total emitted GeV energy vs. duration of GeV emission of {the novae detected in the GeV range}. The duration is defined by the epoch during which the emission with TS > 4 lasted.  The data illustrated with triangles are taken from \protect\cite{Albert2022}. {The symbols with the square and filled circles correspond to the emission energy of nova FM Cir estimated with the geometric distance and photogeometric distance, respectively.}}
     \label{t-energy}
 \end{figure}

{According to GAIA archive\footnote{ \url{https://dc.g-vo.org/gedr3dist/q/cone/form}}, the 
  distance of nova FM Cir is estimated as $3.23^{+0.84}_{-0.54}$~kpc or $7.48^{+1.07}_{-3.05}$~kpc in geometric or photogeometric measurements~\citep{bailer2021}}. To estimate the total emitted energy in the $\gamma$-ray bands, we integrated the daily flux detected with  $TS>4$.  Figure~\ref{t-energy} compares the total emitted energy and duration of the  gamma-ray emission of the novae detected in the GeV bands. We can see in the figure that  the duration of FM Cir  can be the longest among those of $Fermi$-LAT novae, while 
its total emission  is consistent with the others.

\cite{Li2017} reported a strong correlation between the optical and gamma-ray light curves in nova ASASSN-16ma, in which the $\gamma$-ray peak 
aligned with the optical peak, suggesting the optical emission of ASASSN-16ma was a reprocessing  of the shock emission rather than that of the emission from the hot white dwarf. For FM Cir, the correlation analysis by utilizing the discrete correlation function indicated a lag of about 10 days between the gamma-ray peak and 
the optical peak. There is the second peak of the $\gamma$-ray emission at 163 day after the nova eruption. For FM Cir, hence,  we could not find the correlation between the optical and gamma-ray emission with the current data. 

\subsection{X-ray emission properties}
\begin{figure}
    \centering
    \includegraphics[scale=0.38]{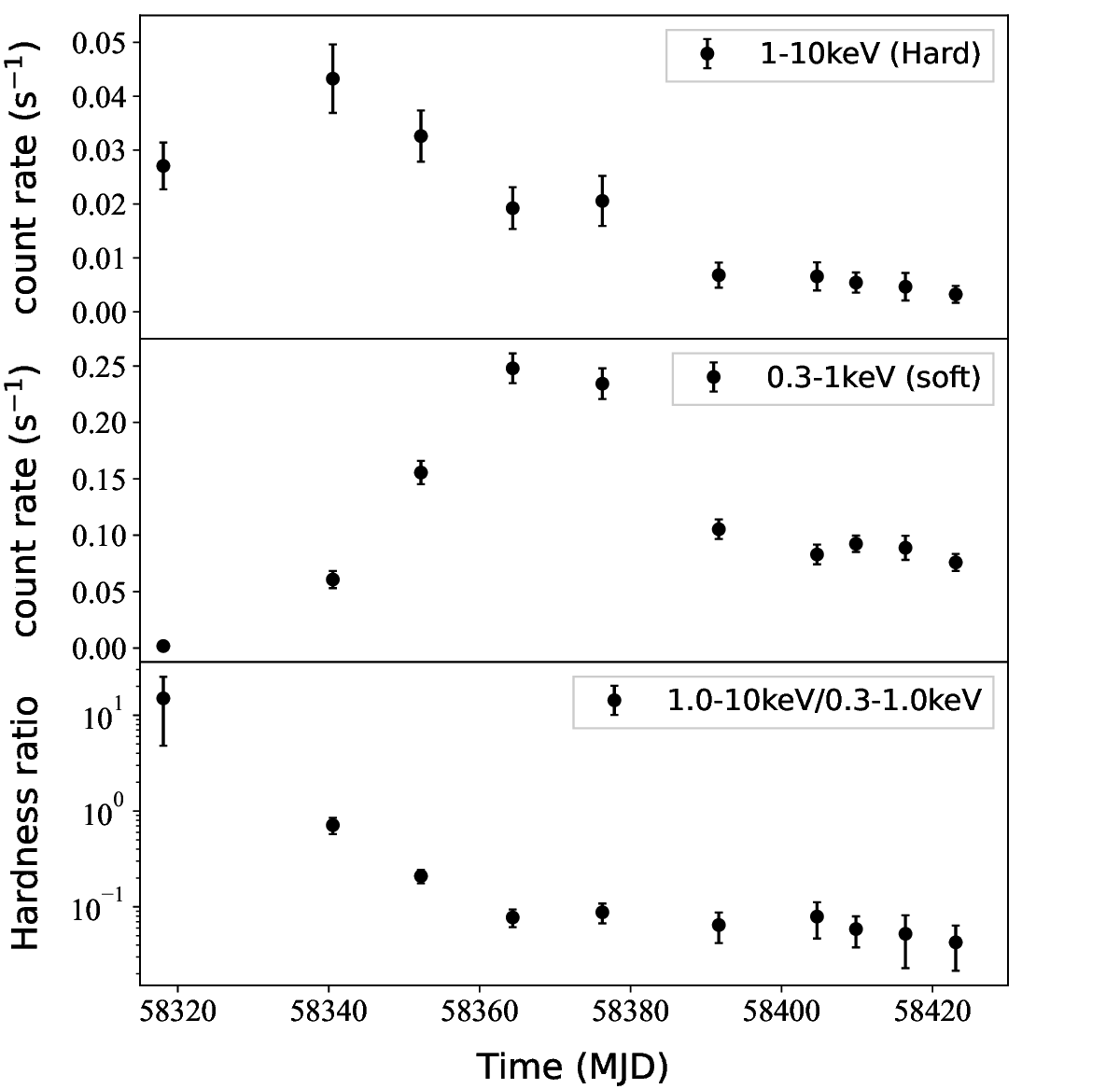}
    \caption{The light curves of hard X-rays (top panel) and soft X-rays (middle panel). The bottom panel shows the hardness ratio (1.0-10.0~keV/0.3-1.0~keV).}
    \label{hardness}
\end{figure}
As Figure~\ref{lighcurve-1d} shows, the significant detection of the X-ray emission of FM Cir started around the epoch when the detection of the $\gamma$-ray emission with $TS>4$ was ended. This behavior of emergence of the X-ray emission after the $\gamma$-ray emission is similar to those of other {novae detected in the GeV range} \citep{Gordon2021}. We can also see that the epoch of the emergence of the X-ray emission coincided with the end of the dust absorption feature in the optical light around MJD~58300.  
These multi-wavelength observations suggest that absorption causes the absence of X-rays after the nova eruption until the environment becomes transparent. 

We fit the observed spectrum with the blackbody radiation (\verb|bbodyrad| in \verb|Xspec|) and/or the optically thin plasma emission (\verb|mekal|) models, as shown in  Table~\ref{table-model}.   The X-ray spectrum just after its emergence  (top panel of Figure~\ref{spectrum-X-ray}) can be fitted by  {a }\verb|mekal|~{ model with a temperature of $k_BT_{mekal}\sim 3$~keV~(Table~\ref{table-model})}, where $k_B$ is the Boltzmann constant.
It may be reasonable to assume that the component of the optically thin thermal emission  originates from the shocked heated plasma, whose  temperature may be  described as ~\citep{Steinberg2018MNRAS}: $k_{B}T \approx 1.2~{\rm keV}~\left(\frac{v}{10^3~{\text{km~s}^{-1}}}\right)^{2},$ where  $v$ is the shock velocity. After the emergence of the X-ray emission, the hardness ratio rapidly decreased (bottom panel in Figure~\ref{hardness}) and the second emission component appears as Figure~\ref{spectrum-X-ray} indicates.

We can see in  Table~\ref{table-model} that the fitting column density $N_H$  tends the decreasing.
After the peak in the observed count rate  (Figure~\ref{hardness}), on the other hand, $N_H$ is not well constrained.  We, therefore, fix the column density to $N_H=2.94\times 10^{21}~{\rm cm^{-2}}$ estimated from the sky position~\footnote{\url{https://heasarc.gsfc.nasa.gov/cgi-bin/Tools/w3nh/w3nh.pl}}. It is found   that the temperature of the blackbody
radiation of the soft component ($k_BT_{BB}$ in Table~\ref{table-model}) is  $k_{B}T_{BB} \sim$ 20-40\, eV,  suggesting the
emission from the hot white dwarf surface; the effective radius at the initial stage of the observed X-ray emission is  of  the order of $10^8$~cm.

As Figure~\ref{lcnova} indicates, the emission was not detected in the single Swift observation before MJD~58300.
If the all data taken before MJD~58300 are stacked, on the other hand, the emission can be detected with a significance level of $4\sigma$. The time averaged
count rate is estimated to be $(2.34\pm0.53)\times 10^{-3}$~count/s, which is about one order of magnitude lower
than $(2.75\pm0.43)\times 10^{-2}$~count/s of the first detection after MJD~58300. This may indicate that the X-ray emission is significantly absorbed before MJD~58300; we could not
fit the observed spectrum  because of the  low count rates. 
  
  For   the  navae from  which  the X-ray emission was detected during the gamma-ray detection, it has been observed
  that the ratio of the GeV luminosity over
  the X-ray luminosity is $L_{\gamma}/L_{X}>10-10^2$, and such a high ratio could be explained by several  emission senarios, for example, multiple shocks, very high efficiency of the acceleration process, etc.,~\citep{Gordon2021}.
  As we described in the previous paragraph, the time averaged count rate before MJD~58300 is about one order of magnitude  smaller than that of the first significant detection with a single Swift  observation. If we just scale the unabsorbed flux using  the count rate,  we obtain the ratio as  $L_{\gamma}/L_{X}\sim 10$. We may, however, expect that the low count rate is due to a stronger absorption, and the time averaged unabsorbed flux is of the order of that of the first detection after MJD~58300. In such a case, we estimate as $L_{\gamma}/L_{X}\sim 1$.  The current data, therefore, could not  provide a  tight constraint on the $L_{\gamma}/L_{X}$. 

\section{conclusion}
We presented an analysis of $Fermi$-LAT observations of nova FM Cir and identified evidence of $\gamma$-ray emission. The detection  significance could exceed $>4\sigma$ for 20-day bins and reached  $\sim 3\sigma$ for daily bins in the data taken at $\sim$ 40 days after the eruption.  We found that  the $\gamma$-ray emission with a TS value larger than $4$
lasted  about 180 days, which is the longest among the known novae detected in the GeV range. On the other hand, the integrated energy  flux  $\sim 10^{41-42}$\,erg was consistent with those of the novae detected in the GeV range.
The X-ray emission emerged at MJD~58318 after the GeV emission was ended.  The X-ray spectrum just after the emergence was fitted by the emission from the optically thin thermal plasma, which likely originates from the shock. 
It was observed that the hardness ratio rapidly decreased with time, and the spectra taken later time were dominated by the blackbody radiation from the hot white dwarf. The multi-wavelength properties in the optical to GeV bands are similar to those of other {novae detected in the GeV range, suggesting that nova FM Cir is likely a nova in the GeV range}.

\section*{Acknowledgements}
We express our appreciation to an anonymous referee for useful comments and suggestions. We acknowledge with thanks the variable star observations from the AAVSO International Database contributed by observers worldwide and used in this research. This work made use of data supplied by the UK Swift Science Data Centre at the University of Leicester.
H.H.W. is supported by the Scientific Research Foundation of Hunan Provincial Education Department (21C0343). J.T. is  supported by the National Key Research and Development Program of China (grant No. 2020YFC2201400) and the National Natural Science Foundation of
China (grant No. 12173014). L.C.-C.L. is supported by NSTC through grants  110-2112-M-006-006-MY3 and 112-2811-M-006-019. P.H.T. is supported by the National Natural Science Foundation of China (NSFC) grant 12273122 and a science research grant from the China Manned Space Project (No. CMS-CSST-2021-B11).
\section*{Data Availability}
(i) The Fermi-LAT data used in this article are available in the LAT data server at https://fermi.gsfc.nasa.gov/ssc/data/ access/. \\
(ii) The X-ray data underlying this paper are available in the Swift
archive at https://www.swift.ac.uk/swift\_live/. \\
(iii) We agree to share data derived in this article on reasonable request to the corresponding author.
\bibliographystyle{mnras}
\bibliography{sample}

\begin{thebibliography}{}
\makeatletter
\relax
\def\mn@urlcharsother{\let\do\@makeother \do\$\do\&\do\#\do\^\do\_\do\%\do\~}
\def\mn@doi{\begingroup\mn@urlcharsother \@ifnextchar [ {\mn@doi@}
  {\mn@doi@[]}}
\def\mn@doi@[#1]#2{\def\@tempa{#1}\ifx\@tempa\@empty \href
  {http://dx.doi.org/#2} {doi:#2}\else \href {http://dx.doi.org/#2} {#1}\fi
  \endgroup}
\def\mn@eprint#1#2{\mn@eprint@#1:#2::\@nil}
\def\mn@eprint@arXiv#1{\href {http://arxiv.org/abs/#1} {{\tt arXiv:#1}}}
\def\mn@eprint@dblp#1{\href {http://dblp.uni-trier.de/rec/bibtex/#1.xml}
  {dblp:#1}}
\def\mn@eprint@#1:#2:#3:#4\@nil{\def\@tempa {#1}\def\@tempb {#2}\def\@tempc
  {#3}\ifx \@tempc \@empty \let \@tempc \@tempb \let \@tempb \@tempa \fi \ifx
  \@tempb \@empty \def\@tempb {arXiv}\fi \@ifundefined
  {mn@eprint@\@tempb}{\@tempb:\@tempc}{\expandafter \expandafter \csname
  mn@eprint@\@tempb\endcsname \expandafter{\@tempc}}}

\bibitem[\protect\citeauthoryear{{Abdo} et~al.,}{{Abdo}
  et~al.}{2010}]{Abdo2010}
{Abdo} A.~A.,  et~al., 2010, \mn@doi [Science] {10.1126/science.1192537}, \href
  {https://ui.adsabs.harvard.edu/abs/2010Sci...329..817A} {329, 817}

\bibitem[\protect\citeauthoryear{{Abdollahi} et~al.,}{{Abdollahi}
  et~al.}{2020}]{FERMI2020}
{Abdollahi} S.,  et~al., 2020, \mn@doi [\apjs] {10.3847/1538-4365/ab6bcb},
  \href {https://ui.adsabs.harvard.edu/abs/2020ApJS..247...33A} {247, 33}

\bibitem[\protect\citeauthoryear{{Abdollahi} et~al.,}{{Abdollahi}
  et~al.}{2022}]{4fgl-dr3}
{Abdollahi} S.,  et~al., 2022, \mn@doi [\apjs] {10.3847/1538-4365/ac6751},
  \href {https://ui.adsabs.harvard.edu/abs/2022ApJS..260...53A} {260, 53}

\bibitem[\protect\citeauthoryear{{Acciari} et~al.,}{{Acciari}
  et~al.}{2022}]{Acciari2022NatAs}
{Acciari} V.~A.,  et~al., 2022, \mn@doi [Nature Astronomy]
  {10.1038/s41550-022-01640-z}, \href
  {https://ui.adsabs.harvard.edu/abs/2022NatAs...6..689A} {6, 689}

\bibitem[\protect\citeauthoryear{{Ackermann} et~al.,}{{Ackermann}
  et~al.}{2014}]{Ackermann2014}
{Ackermann} M.,  et~al., 2014, \mn@doi [Science] {10.1126/science.1253947},
  \href {https://ui.adsabs.harvard.edu/abs/2014Sci...345..554A} {345, 554}

\bibitem[\protect\citeauthoryear{{Albert} et~al.,}{{Albert}
  et~al.}{2022}]{Albert2022}
{Albert} A.,  et~al., 2022, \mn@doi [\apj] {10.3847/1538-4357/ac966a}, \href
  {https://ui.adsabs.harvard.edu/abs/2022ApJ...940..141A} {940, 141}

\bibitem[\protect\citeauthoryear{{Aydi} et~al.,}{{Aydi}
  et~al.}{2020a}]{Aydi2020}
{Aydi} E.,  et~al., 2020a, \mn@doi [Nature Astronomy]
  {10.1038/s41550-020-1070-y}, \href
  {https://ui.adsabs.harvard.edu/abs/2020NatAs...4..776A} {4, 776}

\bibitem[\protect\citeauthoryear{{Aydi} et~al.,}{{Aydi}
  et~al.}{2020b}]{Aydi2020ApJ}
{Aydi} E.,  et~al., 2020b, \mn@doi [\apj] {10.3847/1538-4357/abc3bb}, \href
  {https://ui.adsabs.harvard.edu/abs/2020ApJ...905...62A} {905, 62}

\bibitem[\protect\citeauthoryear{{Bailer-Jones}, {Rybizki}, {Fouesneau},
  {Demleitner}  \& {Andrae}}{{Bailer-Jones} et~al.}{2021}]{bailer2021}
{Bailer-Jones} C.~A.~L.,  {Rybizki} J.,  {Fouesneau} M.,  {Demleitner} M.,
  {Andrae} R.,  2021, \mn@doi [\aj] {10.3847/1538-3881/abd806}, \href
  {https://ui.adsabs.harvard.edu/abs/2021AJ....161..147B} {161, 147}

\bibitem[\protect\citeauthoryear{{Ballet}, {Bruel}, {Burnett}, {Lott}  \& {The
  Fermi-LAT collaboration}}{{Ballet} et~al.}{2023}]{4fgl-dr4}
{Ballet} J.,  {Bruel} P.,  {Burnett} T.~H.,  {Lott} B.,   {The Fermi-LAT
  collaboration} 2023, \mn@doi [arXiv e-prints] {10.48550/arXiv.2307.12546},
  \href {https://ui.adsabs.harvard.edu/abs/2023arXiv230712546B} {p.
  arXiv:2307.12546}

\bibitem[\protect\citeauthoryear{{Chomiuk} et~al.,}{{Chomiuk}
  et~al.}{2014}]{Chomiuk2014Natur}
{Chomiuk} L.,  et~al., 2014, \mn@doi [\nat] {10.1038/nature13773}, \href
  {https://ui.adsabs.harvard.edu/abs/2014Natur.514..339C} {514, 339}

\bibitem[\protect\citeauthoryear{{Chomiuk}, {Metzger}  \& {Shen}}{{Chomiuk}
  et~al.}{2021}]{Chomiuk2021}
{Chomiuk} L.,  {Metzger} B.~D.,   {Shen} K.~J.,  2021, \mn@doi [\araa]
  {10.1146/annurev-astro-112420-114502}, \href
  {https://ui.adsabs.harvard.edu/abs/2021ARA&A..59..391C} {59, 391}

\bibitem[\protect\citeauthoryear{{Della Valle} \& {Izzo}}{{Della Valle} \&
  {Izzo}}{2020}]{Della2020A&AR}
{Della Valle} M.,  {Izzo} L.,  2020, \mn@doi [\aapr]
  {10.1007/s00159-020-0124-6}, \href
  {https://ui.adsabs.harvard.edu/abs/2020A&ARv..28....3D} {28, 3}

\bibitem[\protect\citeauthoryear{{Evans} et~al.,}{{Evans}
  et~al.}{2007}]{Evans2007A&A}
{Evans} P.~A.,  et~al., 2007, \mn@doi [\aap] {10.1051/0004-6361:20077530},
  \href {https://ui.adsabs.harvard.edu/abs/2007A&A...469..379E} {469, 379}

\bibitem[\protect\citeauthoryear{{Evans} et~al.,}{{Evans}
  et~al.}{2009}]{Evans2009MNRAS}
{Evans} P.~A.,  et~al., 2009, \mn@doi [\mnras]
  {10.1111/j.1365-2966.2009.14913.x}, \href
  {https://ui.adsabs.harvard.edu/abs/2009MNRAS.397.1177E} {397, 1177}

\bibitem[\protect\citeauthoryear{{Franckowiak}, {Jean}, {Wood}, {Cheung}  \&
  {Buson}}{{Franckowiak} et~al.}{2018}]{Franckowiak2018}
{Franckowiak} A.,  {Jean} P.,  {Wood} M.,  {Cheung} C.~C.,   {Buson} S.,  2018,
  \mn@doi [\aap] {10.1051/0004-6361/201731516}, \href
  {https://ui.adsabs.harvard.edu/abs/2018A&A...609A.120F} {609, A120}

\bibitem[\protect\citeauthoryear{{Gallagher} \& {Starrfield}}{{Gallagher} \&
  {Starrfield}}{1978}]{Gallagher1978}
{Gallagher} J.~S.,  {Starrfield} S.,  1978, \mn@doi [\araa]
  {10.1146/annurev.aa.16.090178.001131}, \href
  {https://ui.adsabs.harvard.edu/abs/1978ARA&A..16..171G} {16, 171}

\bibitem[\protect\citeauthoryear{{Gordon}, {Aydi}, {Page}, {Li}, {Chomiuk},
  {Sokolovsky}, {Mukai}  \& {Seitz}}{{Gordon} et~al.}{2021}]{Gordon2021}
{Gordon} A.~C.,  {Aydi} E.,  {Page} K.~L.,  {Li} K.-L.,  {Chomiuk} L.,
  {Sokolovsky} K.~V.,  {Mukai} K.,   {Seitz} J.,  2021, \mn@doi [\apj]
  {10.3847/1538-4357/abe547}, \href
  {https://ui.adsabs.harvard.edu/abs/2021ApJ...910..134G} {910, 134}

\bibitem[\protect\citeauthoryear{{H.~E.~S.~S. Collaboration}
  et~al.,}{{H.~E.~S.~S. Collaboration} et~al.}{2022}]{HESS2022}
{H.~E.~S.~S. Collaboration} et~al., 2022, \mn@doi [Science]
  {10.1126/science.abn0567}, \href
  {https://ui.adsabs.harvard.edu/abs/2022Sci...376...77H} {376, 77}

\bibitem[\protect\citeauthoryear{{Li} et~al.,}{{Li} et~al.}{2017}]{Li2017}
{Li} K.-L.,  et~al., 2017, \mn@doi [Nature Astronomy]
  {10.1038/s41550-017-0222-1}, \href
  {https://ui.adsabs.harvard.edu/abs/2017NatAs...1..697L} {1, 697}

\bibitem[\protect\citeauthoryear{{Metzger}, {Hasco{\"e}t}, {Vurm},
  {Beloborodov}, {Chomiuk}, {Sokoloski}  \& {Nelson}}{{Metzger}
  et~al.}{2014}]{Metzger2014}
{Metzger} B.~D.,  {Hasco{\"e}t} R.,  {Vurm} I.,  {Beloborodov} A.~M.,
  {Chomiuk} L.,  {Sokoloski} J.~L.,   {Nelson} T.,  2014, \mn@doi [\mnras]
  {10.1093/mnras/stu844}, \href
  {https://ui.adsabs.harvard.edu/abs/2014MNRAS.442..713M} {442, 713}

\bibitem[\protect\citeauthoryear{{Metzger}, {Finzell}, {Vurm}, {Hasco{\"e}t},
  {Beloborodov}  \& {Chomiuk}}{{Metzger} et~al.}{2015}]{Metzger2015}
{Metzger} B.~D.,  {Finzell} T.,  {Vurm} I.,  {Hasco{\"e}t} R.,  {Beloborodov}
  A.~M.,   {Chomiuk} L.,  2015, \mn@doi [\mnras] {10.1093/mnras/stv742}, \href
  {https://ui.adsabs.harvard.edu/abs/2015MNRAS.450.2739M} {450, 2739}

\bibitem[\protect\citeauthoryear{{Molaro}, {Izzo}, {Bonifacio}, {Hernanz},
  {Selvelli}  \& {della Valle}}{{Molaro} et~al.}{2020}]{Molaro2020MNRAS}
{Molaro} P.,  {Izzo} L.,  {Bonifacio} P.,  {Hernanz} M.,  {Selvelli} P.,
  {della Valle} M.,  2020, \mn@doi [\mnras] {10.1093/mnras/stz3587}, \href
  {https://ui.adsabs.harvard.edu/abs/2020MNRAS.492.4975M} {492, 4975}

\bibitem[\protect\citeauthoryear{{Mukai}, {Orio}  \& {Della Valle}}{{Mukai}
  et~al.}{2008}]{Mukai2008}
{Mukai} K.,  {Orio} M.,   {Della Valle} M.,  2008, \mn@doi [\apj]
  {10.1086/529362}, \href
  {https://ui.adsabs.harvard.edu/abs/2008ApJ...677.1248M} {677, 1248}

\bibitem[\protect\citeauthoryear{{Orio}, {Covington}  \& {{\"O}gelman}}{{Orio}
  et~al.}{2001}]{Orio2001A&A}
{Orio} M.,  {Covington} J.,   {{\"O}gelman} H.,  2001, \mn@doi [\aap]
  {10.1051/0004-6361:20010537}, \href
  {https://ui.adsabs.harvard.edu/abs/2001A&A...373..542O} {373, 542}

\bibitem[\protect\citeauthoryear{{Page}, {Beardmore}  \& {Osborne}}{{Page}
  et~al.}{2020}]{Page2020AdSpR}
{Page} K.~L.,  {Beardmore} A.~P.,   {Osborne} J.~P.,  2020, \mn@doi [Advances
  in Space Research] {10.1016/j.asr.2019.08.003}, \href
  {https://ui.adsabs.harvard.edu/abs/2020AdSpR..66.1169P} {66, 1169}

\bibitem[\protect\citeauthoryear{{Schaefer}}{{Schaefer}}{2021}]{Schaefer2021}
{Schaefer} B.~E.,  2021, \mn@doi [Research Notes of the American Astronomical
  Society] {10.3847/2515-5172/ac0d5b}, \href
  {https://ui.adsabs.harvard.edu/abs/2021RNAAS...5..150S} {5, 150}

\bibitem[\protect\citeauthoryear{{Steinberg} \& {Metzger}}{{Steinberg} \&
  {Metzger}}{2018}]{Steinberg2018MNRAS}
{Steinberg} E.,  {Metzger} B.~D.,  2018, \mn@doi [\mnras]
  {10.1093/mnras/sty1641}, \href
  {https://ui.adsabs.harvard.edu/abs/2018MNRAS.479..687S} {479, 687}

\bibitem[\protect\citeauthoryear{{Strader}, {Chomiuk}, {Swihart}  \&
  {Shishkovsky}}{{Strader} et~al.}{2018}]{Strader2018}
{Strader} J.,  {Chomiuk} L.,  {Swihart} S.,   {Shishkovsky} L.,  2018, The
  Astronomer's Telegram, \href
  {https://ui.adsabs.harvard.edu/abs/2018ATel11209....1S} {11209, 1}

\bibitem[\protect\citeauthoryear{{Vurm} \& {Metzger}}{{Vurm} \&
  {Metzger}}{2018}]{vurm2018}
{Vurm} I.,  {Metzger} B.~D.,  2018, \mn@doi [\apj] {10.3847/1538-4357/aa9c4a},
  \href {https://ui.adsabs.harvard.edu/abs/2018ApJ...852...62V} {852, 62}

\makeatother
\end{thebibliography}

\appendix
\section{X-ray spectrum of FM Cir}
\label{appendix}
\begin{figure}
    \includegraphics[scale=0.28,angle=270]{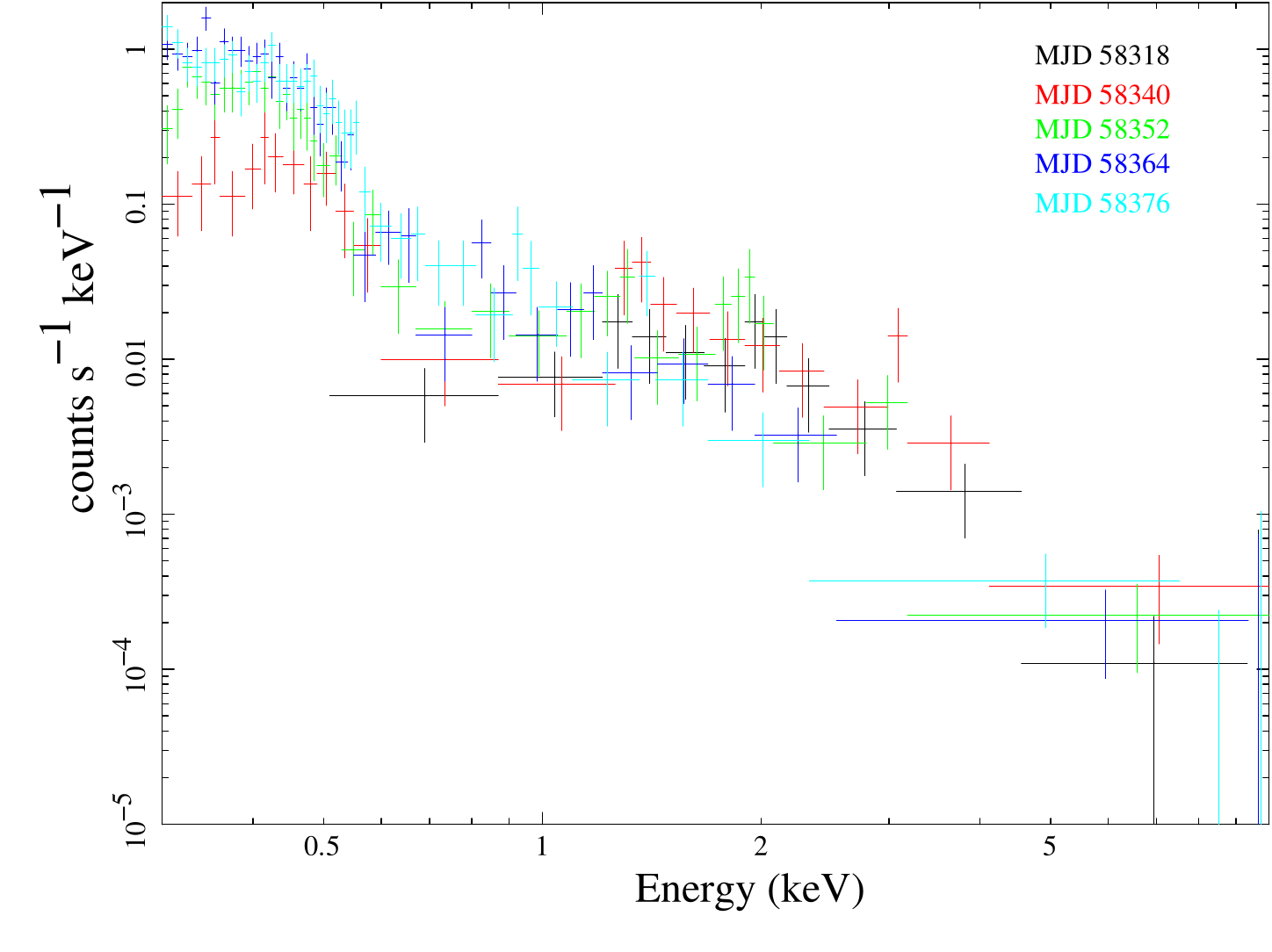}
\includegraphics[scale=0.28,angle=270]{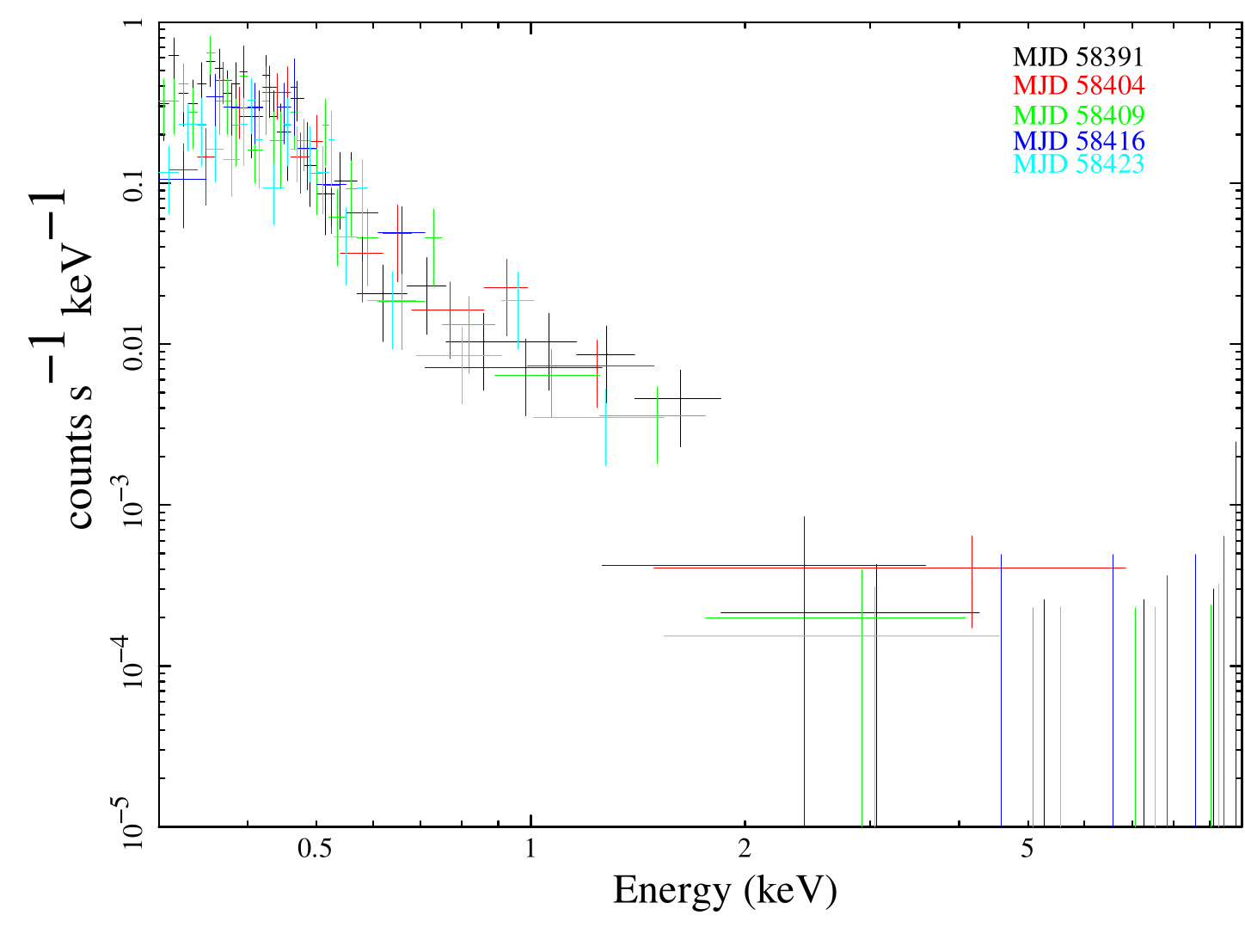}
    \caption{The spectra of nova FM Cir from $Swift$-XRT data in 0.3-10.0~keV.}
    \label{spectrum-X-ray-all}
\end{figure}

{Figure~\ref{spectrum-X-ray-all} shows the temporal  evolution of the  X-ray spectra taken by  $Swift$-XRT. 
In the initial observation (black color in the top panel) did not show soft blackbody radiation, which is likely produced by 
the surface emission of the white dwarf. The observed emission reached the peak flux around MJD~58364.}

\end{document}